\documentclass{sf2a-conf2015}
\usepackage{graphicx}
\usepackage{hyperref}
\usepackage[]{natbib}  
\usepackage{epstopdf}
\usepackage{amssymb}
\newcommand{\um}{\,$\mu$m\,}

\newcommand{\cs}{C$_{60}$\,}
\newcommand{\csp}{C$_{60}^+$}

\def\BibTeX{{\rm B\kern-.05em{\sc i\kern-.025em b}\kern-.08em
    T\kern-.1667em\lower.7ex\hbox{E}\kern-.125emX}}
\bibpunct{(}{)}{;}{a}{}{,}  

\begin{document}

\TitreGlobal{SF2A 2015}


\title{30 years of cosmic fullerenes}

\runningtitle{30 yrs of Cosmic fullerenes}

\author{O. Bern\'e}\address{IRAP Universit\'e de Toulouse; UPS-OMP; IRAP;  Toulouse, France
 CNRS; 9 Av. colonel Roche, BP 44346, F-31028 Toulouse cedex 4, France}

\author{J. Montillaud}\address{Institut Utinam, CNRS UMR 6213, OSU THETA, Universit\'e de Franche-Comt\'e, 41bis avenue de l'Observatoire, 25000 Besan\c{c}on, France}

\author{G. Mulas}\address{Istituto Nazionale di Astrofisica -- Osservatorio Astronomico di Cagliari -- strada 54, localitˆ Poggio dei Pini, 09012-- Capoterra (CA), Italy}

\author{C. Joblin$^{1}$}




\setcounter{page}{237}


\maketitle


\begin{abstract}
In 1985, ``During experiments aimed at understanding the mechanisms by which long-chain carbon molecules are formed in interstellar space and circumstellar shells'', Harry Kroto and his collaborators serendipitously discovered a new form of carbon: fullerenes. The most emblematic fullerene (i.e. C$_{60}$ ``buckminsterfullerene''), contains exactly 60 carbon atoms organized in a cage-like structure similar to a soccer ball. Since their discovery impacted the field of nanotechnologies, Kroto and colleagues received the Nobel prize in 1996. The cage-like structure, common to all fullerene molecules, gives them unique properties, in particular an extraordinary stability. For this reason and since they were discovered in experiments aimed to reproduce conditions in space, fullerenes were sought after by astronomers for over two decades, and it is only recently that they have been firmly identified by spectroscopy, in evolved stars and in the interstellar medium. This identification offers the opportunity to study the molecular physics of fullerenes in the unique physical conditions provided by space, and to make the link with other large carbonaceous molecules thought to be present in space : polycyclic aromatic hydrocarbons. 
\end{abstract}



\section{The presence of large carbonaceous molecules in space : the PAH hypothesis} 

About 30 years ago, the presence of bands  in emission  (the strongest of  which  are found at 3.3,~6.2,~7.7,~8.6,~11.2 and 12.7 \um)
in the mid-infrared  spectrum of  our Galaxy was observed. Soon after that, it was proposed by \citet{leg84} and then \citet{all85}  
that these bands result from the emission of large carbonaceous molecules from the family of  Polycyclic Aromatic Hydrocarbons (PAHs), 
present in the gas phase and heated  by the absorption of single UV photons. Since then, PAHs have been invoked to be part of 
numerous key processes in  interstellar and circumstellar environments (e.g. the formation of H$_2$, heating of the  neutral gas, UV extinction etc.) 
and are widely used as tracers of physical conditions and star-formation in galaxies. Yet, even though it is widely accepted, 
the gas phase PAH model remains an hypothesis because no specific PAH molecule could be identified up to date. This is mainly 
because there exists a large number of PAH molecules (it is a "family" of molecules) and the broad mid-infrared emission bands are  
 not specific enough to identify individual species from the PAH family.

\begin{figure}[ht!]
 \centering
 \includegraphics[width=0.4\textwidth,clip]{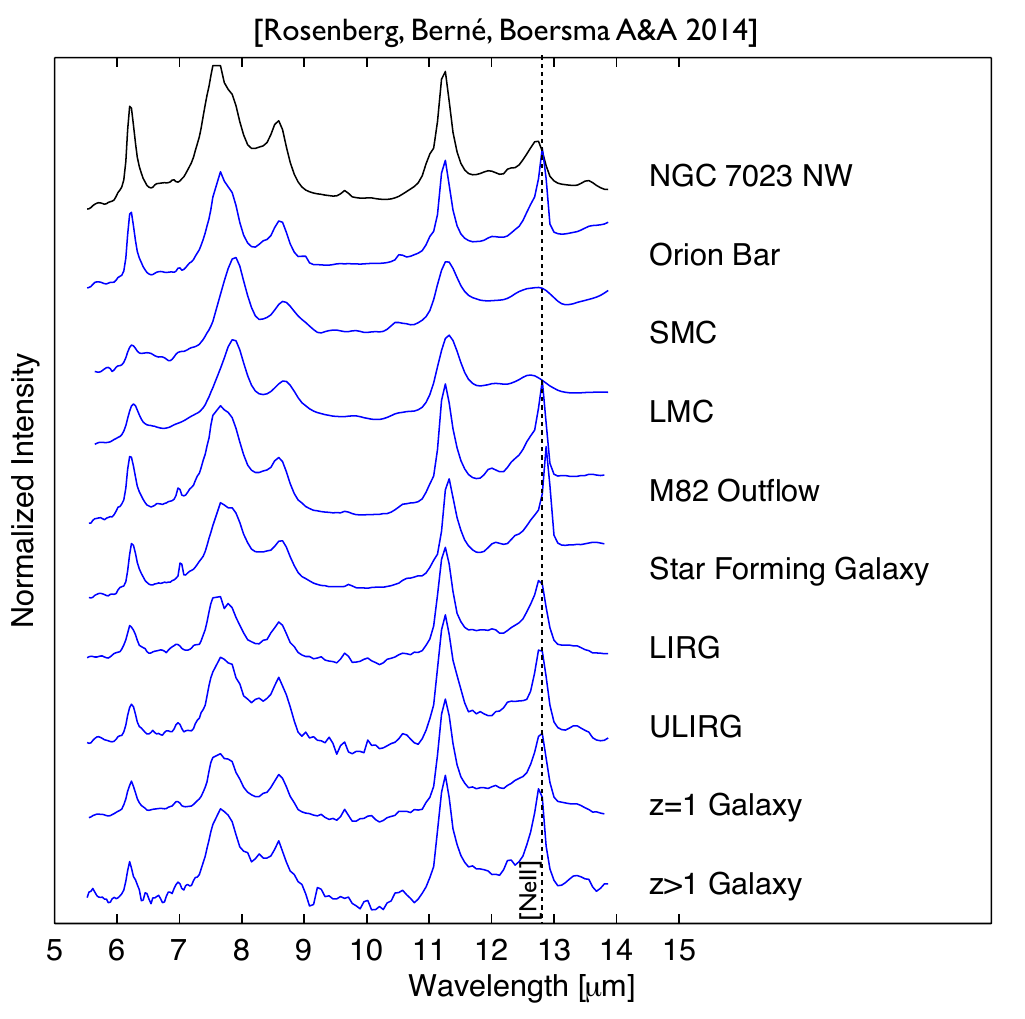}      
  \caption{Mid-infrared spectra of  UV-irradiated regions in the Milky Way and other galaxies obtained with the IRS instrument on- board the Spitzer Space Telescope and showing the main PAH bands. (see \citealt{ros14} for details).}
  \label{fig.1}
\end{figure}

\section{The serendipitous discovery of the \cs molecule}

\begin{figure}[ht!]
 \centering
 \includegraphics[width=0.4\textwidth,clip]{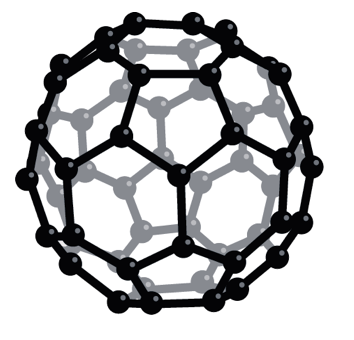}      
  \caption{Structure of the \cs molecule, ``Buckminsterfullerene''. Figure credit L. Cadars. }
  \label{fig.2}
\end{figure}

About at the same time the PAH model was proposed, H. Kroto and his collaborators serendipitously formed, 
in the laboratory, a new molecule made of 60 carbon atoms 
arranged as the vertices of a soccer ball (\cs , Fig.~\ref{fig.2}). It is worth noting that, in the original Nature paper \citep{kro85}, 
the authors of the discovery present a photograph of a soccer ball as their Fig. 1. 
Because of the resemblance of the geometry of this molecule with the buildings of 
architect Buckminster Fuller, they coined the name "Buckminsterfullerene" for this new molecule. In fact, there exists a whole class of 
pure carbon molecules with icosahedral geometries of different sizes and shapes, which belong to the family of fullerenes. 
The discovery of \cs and fullerenes had a strong impact on the development of nanotechnologies, and H. Kroto and his colleagues 
received the Nobel prize in 1996 for this discovery. Yet, it should not be forgotten that, in their 1985 paper, the future Nobel laureates 
start their article by ``During experiments aimed at understanding the mechanisms by which long-chain carbon molecules 
are formed in interstellar space and circumstellar shells'', i.e. that their main goal was to understand interstellar and 
circumstellar chemistry. 

\section{The search and discovery of \cs in space}

\begin{figure}[ht!]
 \centering
 \includegraphics[width=0.5\textwidth,clip]{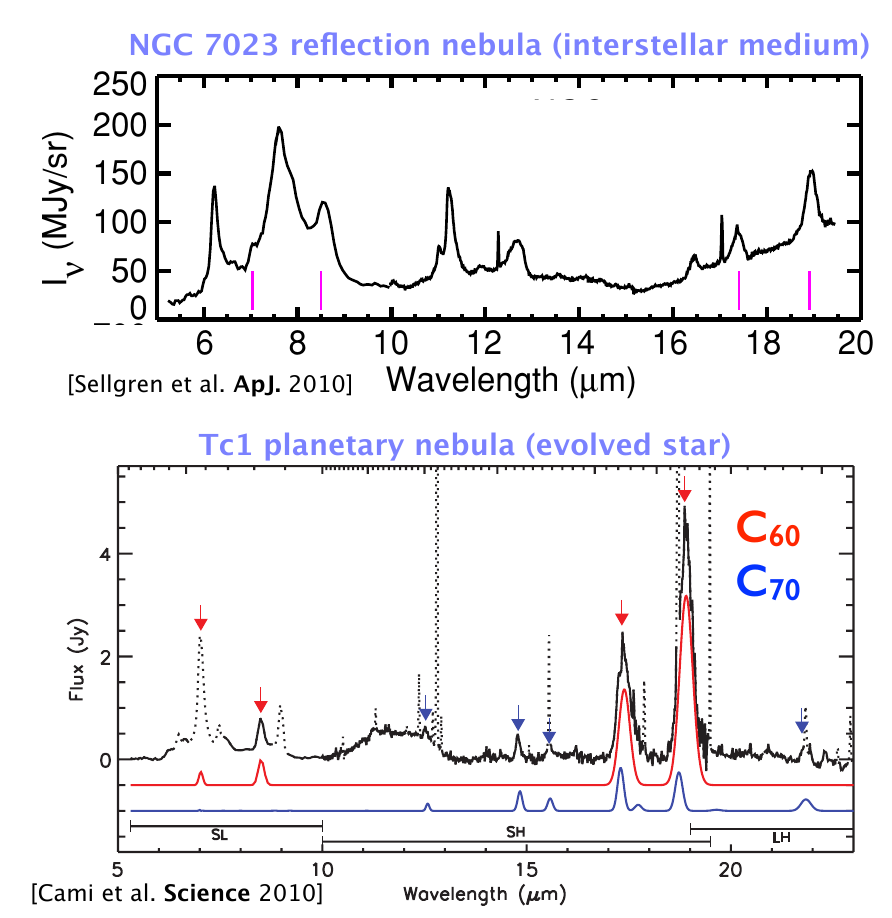}      
  \caption{Spitzer spectra of the NGC 7023 reflection nebula and TC 1 planetary nebula in which the main signatures of 
  the \cs molecule are observed. In the case of NGC 7023, the PAH bands are also present. Figures from \citet{sel10} and 
  \citet{cam10}. }
  \label{fig.3}
\end{figure}

\cs has been sought for by astronomers since its discovery in 1985. The first serious evidence of the presence of \cs, 
in reality its cation i.e. \csp, was given by \citet{foi94} who detected two weak absorption bands in the diffuse interstellar 
medium with wavelengths close to those measured in the laboratory for the electronic transitions of \csp. Unfortunately, 
the laboratory data was obtained using matrix isolation techniques, which do not allow a precise enough measurement
of the band positions for a definitive identification. Therefore, this detection 
remained debated for nearly two decades. It was only in 2010 that clear evidence for the presence of \cs was provided.
This time, it was with the detection in emission of the main vibrational bands 
(Fig.~\ref{fig.3}) of the neutral \cs molecule, in two reflection nebulae by \citet{sel10} and in an evolved star by \citet{cam10}.
While this discovery was an important step, it raised a number of question for astrochemistry :
\begin{itemize}
\item{How is \cs formed in space ?}
\item{\cs was detected in its neutral form, while given the harsh conditions of the ISM one could expect to detect the cationic form of the molecule, i.e. \csp}
\item{It was proposed by \citet{sel10} that \cs is in the gas phase and heated by single UV photons, as in the PAH model, but \citet{cam10} suggested that \cs is stuck on the surface of grains at thermal equilibrium with the radiation field.}
\end{itemize}

\section{The formation of \cs}

\begin{figure}[ht!]
 \centering
 \includegraphics[width=0.8\textwidth,clip]{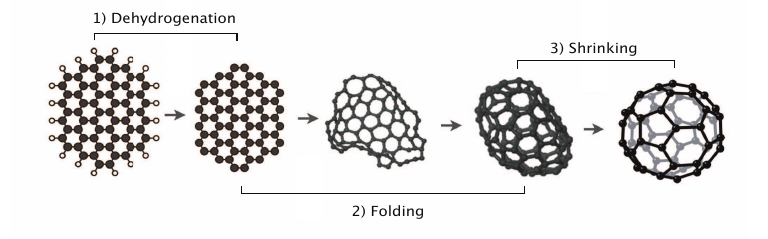}      
  \caption{Schematic representation of the evolutionary scenario for the ``top-down'' formation of fullerenes from PAHs under UV irradiation. From \citet{ber15}.}
  \label{fig.4}
\end{figure}

One question related to fullerenes, and in particular \cs, concerns their formation pathway.
Recently, ``top-down" schemes where larger carbon clusters shrink to reach \cs  have been proposed \citep{chu10, zha13, pie14},
and can be opposed to the traditional ``bottom-up" approach where \cs is built up from small gas-phase species \citep{kro88, hea92, hun94, dun13}.
Using infrared observations of the NGC 7023 nebula, \citet{ber12} found evidence of an increase of the abundance 
of \cs with increasing UV field, while the abundance of PAHs decreases. This was interpreted by these authors as 
evidence for the formation of \cs from large PAHs ($N_C>60$)  under UV irradiation, a top-down mechanism similar
to the one observed by  \citet{chu10}. \citet{her10, her11} and \citet{mic12} proposed a similar mechanism where the starting 
materials are more complex, such as hydrogenated amorphous carbon instead of PAHs. Top-down scenarios are 
particularly appealing, given that the densities prevailing in the ISM are many orders of magnitude too low to allow 
for a bottom-up formation (i.e. starting from small gas-phase species) over reasonable timescales. \citet{ber15} proposed the 
first detailed model of the top-down photochemistry of interstellar fullerenes (Fig.~\ref{fig.4}). 
PAHs are assumed to be formed in the envelopes of evolved stars \citep{fre89, che92, mer14}
and then to be injected in the ISM. According to \citet{ber15}, under UV irradiation, large PAHs, ($60~<~$N$_C \lesssim 1000$) are first fully dehydrogenated 
into small graphene flakes, dehydrogenation being by far the dominant dissociation channel \citep[see][and references 
therein]{mon13}.  Additional UV irradiation{ enables} these flakes to fold into closed cages. Once the cages are closed, they can lose
C$_2$ units if they continue to absorb energy \citep{irl06}.  Because of the low densities prevailing in the 
ISM, the reverse reaction, that is,{ addition} of C$_2$ , is{ too slow to balance 
photodissociation} and therefore the molecule will shrink. Once a system has reached C$_{60}$ , it will remain in this form for a very long time
because it is remarkably stable. \citet{ber15} find that, with this route, it is possible to convert about 1\%
of the interstellar PAHs into \cs. This efficiency results in predicted abundances that are comparable to the observed ones. 
It should be noted that  \cs was recently formed in the laboratory \citep{zhe14} in a top down manner similar to the one
described in this theoretical model, i.e. from the irradiation of PAH molecules in the gas phase. 
However, the detailed steps taken to convert PAHs to \cs in this top down scheme are still a subject of debate within specialists
(see discussion in \citealt{ber15}).

\section{The detection of \csp}

Recently, \citet{ber13} examined in detail the \emph{Spitzer} IRS spectra of the NGC 7023 reflection nebula, at a position close (7.5'') 
to its illuminating B star HD 200775, and found four previously unreported bands at 6.4, 7.1, 8.2, and 10.5 $\mu$m (Fig. 5), in addition to 
the classical bands attributed to PAHs and neutral \cs. These 4 bands are observed 
only in this region of the nebula (Fig. 5), while \cs\, emission is still present slightly farther away from the star, and PAH emission even 
farther away. Based on this observation \citet{ber13} suggested that these bands could be due to \csp. In addition, they conducted 
quantum chemistry calculations to determine the theoretical positions of the \csp\  bands. These theoretical band positions were found to
match very well with the observed ones (Fig. 5). On this basis, \citet{ber13} concluded that the cationic form of \cs, i.e. \csp\  is also 
present in the ISM. In 2015, further evidence for the presence of \csp\  in the ISM was provided by \citet{cam15} who measured in the laboratory
a gas phase electronic spectrum of \csp. The measured positions of the electronic bands were found to be in very good agreement 
with the bands observed by \citet{foi94} (See sect. 3). The detection of an ion,  \csp\ , in emission,  confirms the idea that large carbon 
molecules exist in the gas phase in the ISM  and that their emission is caused by the absorption of individual UV photons
as initially suggested by \citet{sel10}. This brings strong evidence that the PAH model (Sect. 1) is correct.

\begin{figure}[ht!]
 \centering
 \includegraphics[width=0.8\textwidth,clip]{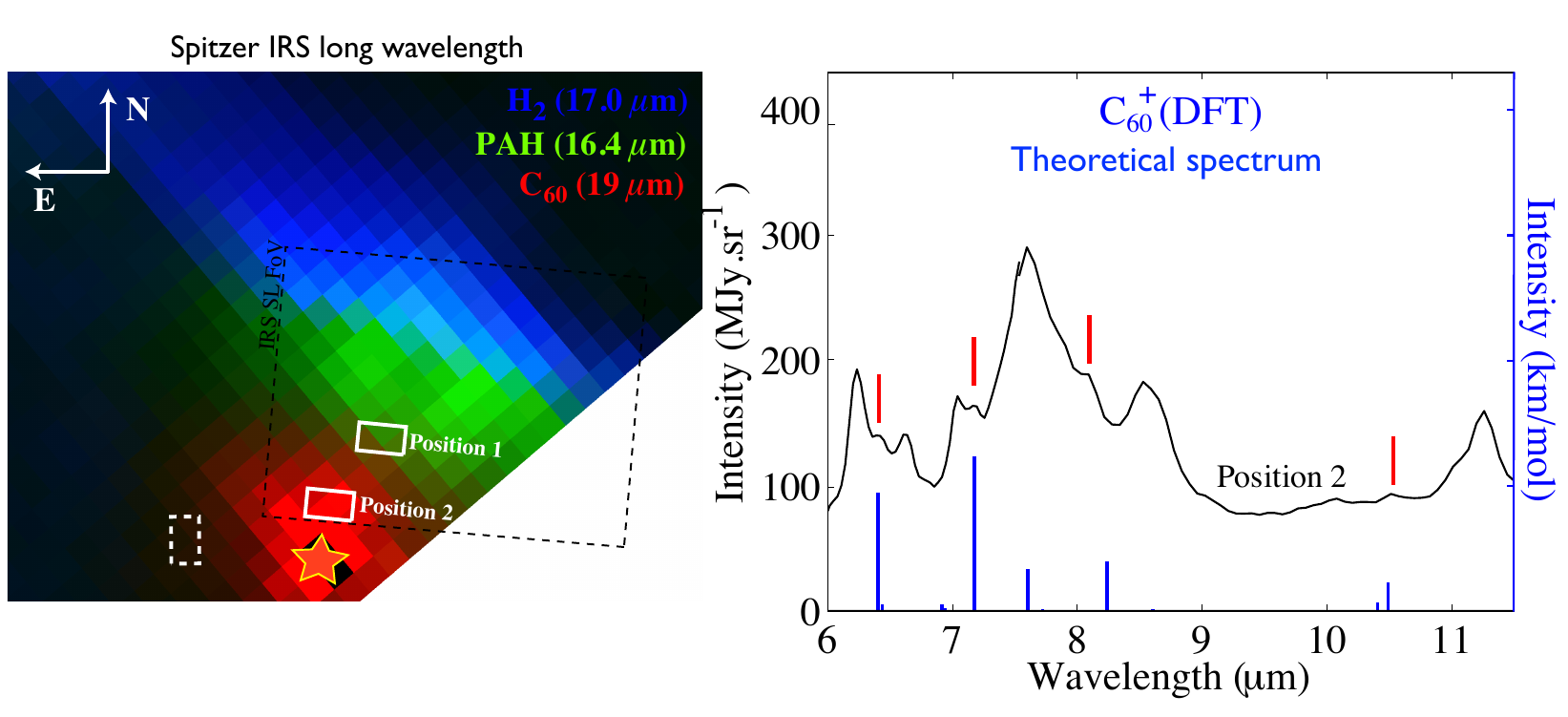}      
  \caption{ \emph{Left:} False-color image of the NGC 7023 nebula such as the one presented in \citet{sel10}, obtained from integrating
different components in the \emph{Spitzer}-IRS  LL spectral cube. Red is the emission integrated in the \cs\, 19$\mu$m band.
Green is the emission of the PAH 16.4 $\mu$m band. Blue is the emission integrated in the H$_2$ (0-0) S(0) 17.0 $\mu$m 
band.  \emph{Right:} Spectrum at positions 1 (see \citealt{ber13} for spectrum at position 2) in the image.
{ Error bars are not shown here but are comparable to the width of the line}. The red lines indicate the four newly detected bands 
attributed to \csp. The DFT calculated spectrum is shown as a bar graph in blue. See \citet{ber13} for details.
\label{Spectra}}
  \label{fig.5}
\end{figure}



\section{Conclusions}
30 years after it was discovered in the laboratory, there is now convincing evidence that \cs and likely other fullerenes are present 
in space (indeed there is evidence for C$_{70}$ in TC1, see Fig.~\ref{fig.3}). In the ISM, \cs is probably formed by a top down mechanism from PAHs, but it is possible that other mechanisms 
are at play in the dense envelopes of evolved stars.
It was shown that \cs  exists in the gas phase in the ISM and is heated by single UV photons emitted by massive stars. 
This confirms the emission mechanism put forward by the PAH model 30 years ago. The presence of neutral and cationic fullerenes 
indicates that the molecule is stable in both states and that the ionization fraction of \cs will depend on the local physical conditions. This
opens the possibility to use \cs and \csp\ emission bands as a proxy of the physical conditions, mainly radiation field which controls the ionization
and gas density which controls the electron recombination rate.

\begin{acknowledgements}
This work was supported by the CNRS program "Physique et Chimie du Milieu Interstellaire" (PCMI).
\end{acknowledgements}

\bibliographystyle{aa}  
\bibliography{berne} 

\end{document}